\documentclass[10pt]{article}

\usepackage{amsmath}
\usepackage{amssymb}

\usepackage{graphicx}

\usepackage{cite}
\usepackage{array}
\usepackage{tabulary}
\usepackage{color} 
\usepackage{wrapfig }

\usepackage[labelfont=bf,labelsep=period,justification=raggedright]{caption}

\makeatletter
\renewcommand{\@biblabel}[1]{\quad#1.}
\makeatother

\date{}

\begin{document}

\begin{flushleft}
{\Large
\textbf{arrayMap: A Reference Resource for Genomic Copy Number Imbalances in Human Malignancies}
}\newline
\\
Haoyang Cai$^{1,\#}$, 
Nitin Kumar$^{1, \#}$, 
Michael Baudis$^{1,\ast}$
\\
\bf{1} Institute of Molecular Life Sciences, University of Zurich, Zurich, Switzerland
\\
$\ast$ E-mail: michael.baudis@imls.uzh.ch\\
\# These authors contributed equally to this work.
\end{flushleft}

\section*{Abstract}
\textbf {\emph {Background:}} The delineation of genomic copy number abnormalities (CNAs) from cancer samples has been instrumental for identification of tumor suppressor genes and oncogenes and proven useful for clinical marker detection. An increasing number of projects have mapped CNAs using high-resolution microarray based techniques. So far, no single resource does provide a global collection of readily accessible oncogenomic array data.\newline

\noindent \textbf {\emph {Methodology/Principal Findings:}} We here present arrayMap, a curated reference database and bioinformatics resource targeting copy number profiling data in human cancer. The arrayMap database provides a platform for meta-analysis and systems level data integration of high-resolution oncogenomic CNA data. To date, the resource incorporates more than 40,000 arrays in 224 cancer types extracted from several resources, including the NCBI's Gene Expression Omnibus (GEO), EBIÕs ArrayExpress (AE), The Cancer Genome Atlas (TCGA), publication supplements and direct submissions. For the majority of the included datasets, probe level and integrated visualization facilitate gene level and genome wide data review. Results from multi-case selections can be connected to downstream data analysis and visualization tools.\newline

\noindent \textbf {\emph {Conclusions/Significance:}} To our knowledge, currently no data source provides an extensive collection of high resolution oncogenomic CNA data which readily could be used for genomic feature mining, across a representative range of cancer entities. arrayMap represents our effort for providing a long term platform for oncogenomic CNA data independent of specific platform considerations or specific project dependence. The online database can be accessed at http://www.arraymap.org.
\section*{Author Summary}

\section*{Introduction}

Genomic copy number abnormalities (CNAs) are a relevant feature in the development of basically all forms of human malignancies \cite{Stallings:2007fh}. Many genomic imbalances are recurrent and display tumor-specific patterns \cite{Myllykangas:2006hv,Weir:2007km}. It is believed that these genomic instabilities reveal mutations in tumor suppressor genes and oncogenes which eventually result in a clone of fully malignant cells. Investigation of CNA hot spots (chromosomal loci frequently involved in CNA) has proven to be an effective methodology to identify novel cancer-causing genes \cite{Wiedemeyer:2008kl,Mullighan:2007jv}. On a systems level, CNA data along with expression or somatic mutation data is used to detect pathways altered in cancers and to deduce functional relevance of pathway members \cite{CancerGenomeAtlasResearchNetwork:2008gr,Kan:2010fo}. Since many CNAs have been attributed to specific tumor types or clinical risk profiles, in some entities copy number profiling is employed to characterize different biological as well as clinical subtypes with implications for treatment and individual prognosis. Subtype-associated CNA regions are used to predict causative genes, furthering understanding of biological differences and leading to discovery of new therapeutic targets \cite{Bergamaschi:2006fj,Hu:2009ez}.

Throughout the last two decades, molecular-cytogenetic techniques have been applied to scan genomic copy number profiles in virtually all types of human neoplasias. For whole genome analysis, these techniques predominantly consist of chromosomal and array comparative genomic hybridization (CGH), including CNA detection by cDNA and single nucleotide polymorphism (SNP) arrays \cite{Kallioniemi:1992ud,Pollack:1999by,Bignell:2004bu} \footnote{In this article, we use the terms "array CGH" and "aCGH" for all technical variants of whole genome copy number arrays. This includes e.g. single color arrays for which regional copy number normalization is performed through bioinformatics procedures applied to external references and internal data distribution.}. While chromosomal CGH has a limited spatial resolution of several megabases, the resolution of recent array based technologies (aCGH) is mainly limited due to cost/benefit evaluations instead of technical obstacles. 

The flood of new insights into structural genomic changes in health and disease has led to an increased interest in genomic data sets in genetic and cancer research. Several systematic studies of CNAs across many cancer types have been performed \cite{Baudis:2007er,Alloza:2011be}. These efforts attempt a more complete understanding of functional effect of CNAs in the context of cancer.

The exponential increase of high resolution CNA datasets offers new challenges and opportunities for large-scale genomic data mining, data modeling and functional data integration. Several online resources have been developed, focusing on different aspects of data content as well as representation \cite{Barrett:2011gr,Parkinson:2010fj,CancerGenomeAtlasResearchNetwork:2008gr,Scheinin:2008ff,Cao:2011ft,Baudis:2001uaa}. An overview of some of the prominent examples is given in Table 1. In principle, these databases facilitate access and utilization of CNA data. However, they are limited to specific aCGH platforms and/or single institutions as well as limited disease categories, or, as in the cases of GEO \cite{Barrett:2011gr} and Ensembl ArrayExpress \cite{Parkinson:2010fj}, mainly serve as raw data repositories. To the best of our knowledge, no single data source does yet provide an extensive collection of high resolution oncogenomic CNA data which readily could be used for genomic feature mining, across a representative range of cancer entities.

Here we present "arrayMap", a web-based reference database for genomic copy number data sets in cancer. We have generated a pipeline to accumulate and process oncogenomic array data into a unified and structured format. The resource incorporates associated histopathological and clinical information where accessible.

So far, arrayMap contains more than 40,000 arrays on 224 cancer types from five main data sources: NCBI GEO, EBI ArrayExpress, The Cancer Genome Atlas, publication supplements and user submitted data. Samples of interest can be browsed, visualized and analyzed via an intuitive interface. Computational tools are provided for biostatistical data analysis such as CNA clustering for case specific or for subset data and basic clinical correlations. arrayMap is publicly available at www.arraymap.org.

\begin{table}[ht]

\caption{Prominent online resources of genomic data}

\setlength{\extrarowheight}{3pt}
\centering
\begin{tabular}{p{2.2cm}p{3.8cm}cp{3.2cm}p{3.2cm}}

\hline
\footnotesize{Name} & \footnotesize{Address} & \footnotesize{Platform(s)} & \footnotesize{Data format} & \footnotesize{Comment}\\
\hline
\footnotesize{GEO \cite{Barrett:2011gr}	} & \footnotesize{ www.ncbi.nlm.nih.gov/geo	} & \footnotesize{ 263	} & \footnotesize{ raw and normalized probe signal intensity	} & \footnotesize{ largest microarray data repository}\\
\footnotesize{ArrayExpress*\cite{Parkinson:2010fj}	} & \footnotesize{ www.ebi.ac.uk/arrayexpress		} & \footnotesize{16} & \footnotesize{ raw and normalized probe signal intensity} & \footnotesize{ many duplicate data in GEO}\\
\footnotesize{TCGA \cite{CancerGenomeAtlasResearchNetwork:2008gr}	} & \footnotesize{ cancergenome.nih.gov} & \footnotesize{1} & \footnotesize{ segmentation data} & \footnotesize{ raw probe data is limited to download}\\
\footnotesize{CanGEM**\cite{Scheinin:2008ff}	} & \footnotesize{www.cangem.org	} & \footnotesize{ 38	} & \footnotesize{normalized probe signal intensity} & \footnotesize{including many types of microarray data}\\
\footnotesize{CaSNP \cite{Cao:2011ft}} & \footnotesize{cistrome.dfci.harvard.edu/CaSNP	} & \footnotesize{ 8	} & \footnotesize{ average copy number and graphic} & \footnotesize{focus on SNP array data}\\
\footnotesize{Progenetix \cite{Baudis:2001uaa}	} & \footnotesize{ www.progenetix.org	} & \footnotesize{ 235	} & \footnotesize{ ISCN*** and golden path	} & \footnotesize{ data from publications}\\
\hline

\end{tabular}

\scriptsize{
\begin{flushleft}Data up to 29 April, 2011\\
*excluding data both in GEO and ArrayExpress\\
** statistical information only including CGH, SNP and cDNA data\\
*** International system for human cytogenetic nomenclature\\
\end{flushleft}
}
\label{tab:table1}

\end{table}


\section*{Results}

\subsection*{Data content}

Our combination of both "top-down" (publication driven) as well as "bottom-up" (array data driven) approaches allowed us to identify a comprehensive set of accessible aCGH based cancer CNA data sets and to estimate the ratio of accessible data of the overall published/deposited data.

As main result of the array data driven approach, we extracted 495 series comprising of 32002 arrays, generated on 237 platforms from NCBIÕs GEO. Among those, raw data files of approximately 29000 whole genome arrays were suitable for inclusion into our data processing pipeline. When reviewing the content of AE, we found that the majority of AE cancer genome data sets were also submitted to GEO. At the time of writing, 11 datasets including 712 arrays not present in GEO had been processed based on AE specific series. Detailed information on the GEO/AE data sets is provided in supplementary Table S1.

The top-down procedure was based on our group's continuous monitoring of cancer related articles utilizing genome copy number screening approaches, as established for our "Progenetix" project (www.progenetix.org; \cite{Baudis:2001uaa}). The census date for the literature based data collection was August 15 2011. At this point, we had identified 931 articles discussing a total of 53213 genomic cancer CNA profiles based on aCGH techniques. Of these, 8728 cases out of 199 articles so far had been extracted from publication related sources (e.g. supplementary data tables) and annotated and made been accessible through Progenetix. This data included cases for which only supervised information but no probe data was available (e.g. author annotated Golden Path or cytogenetic CNA regions). Literature based data sets containing probe specific data or with the respective data presented to us by the authors (640 samples) were included into our arrayMap data processing pipeline.

The data content of arrayMap is summarized in Table 2. Current numbers on the website will include changes based on ongoing  annotation efforts (i.e. addition of data sets, removal of low quality arrays).

As a by-product of our data collection and annotation efforts, we are able to provide estimates of content and trends for the platform usage and cancer entity coverage for the majority of published data. According to the assigned ICD-O 3 (International Classification of Diseases for Oncology, 3rd Edition) code and descriptive diagnostic text, breast carcinoma predominates as single largest clinical entity with 6459 arrays. Supplemental Table S2 presents sample sets in arrayMap classified by ICD-O code.

\begin{table}[!ht]
\caption{aCGH data integrated in arrayMap}
\setlength{\extrarowheight}{4pt}
\centering
\begin{tabular}{llllll}
\hline
Data Source			& Arrays			& Cases	& Series	& Platforms	&Publications\\
\hline
GEO					& 32002			& 25728	& 495	& 237		& 490\\
ArrayExpress			& 712			&		& 11		& 16			& 11\\
TCGA				& 7249			& 3594	& 19		& 1			&*\\
Publication Supplements	& \textgreater4578**	& 4578	&		&			& 137\\
Author Submission		& 556			& 539	& 8		& 7			&\\
\hline
\end{tabular}

\scriptsize{
\begin{flushleft}Data up to 29 April, 2011\\
* Due to lack of publication information, there may be a small amount of duplicate data in GEO\\
**Array number may be higher than case number since reported results per case occasionally may be based on more than one array. The number does not include data presented both in publication supplements as well as GEO.
\end{flushleft}
}
\label{tab:table2}
 \end{table}

\begin{figure}[t]
\centering
\includegraphics[width=5in]{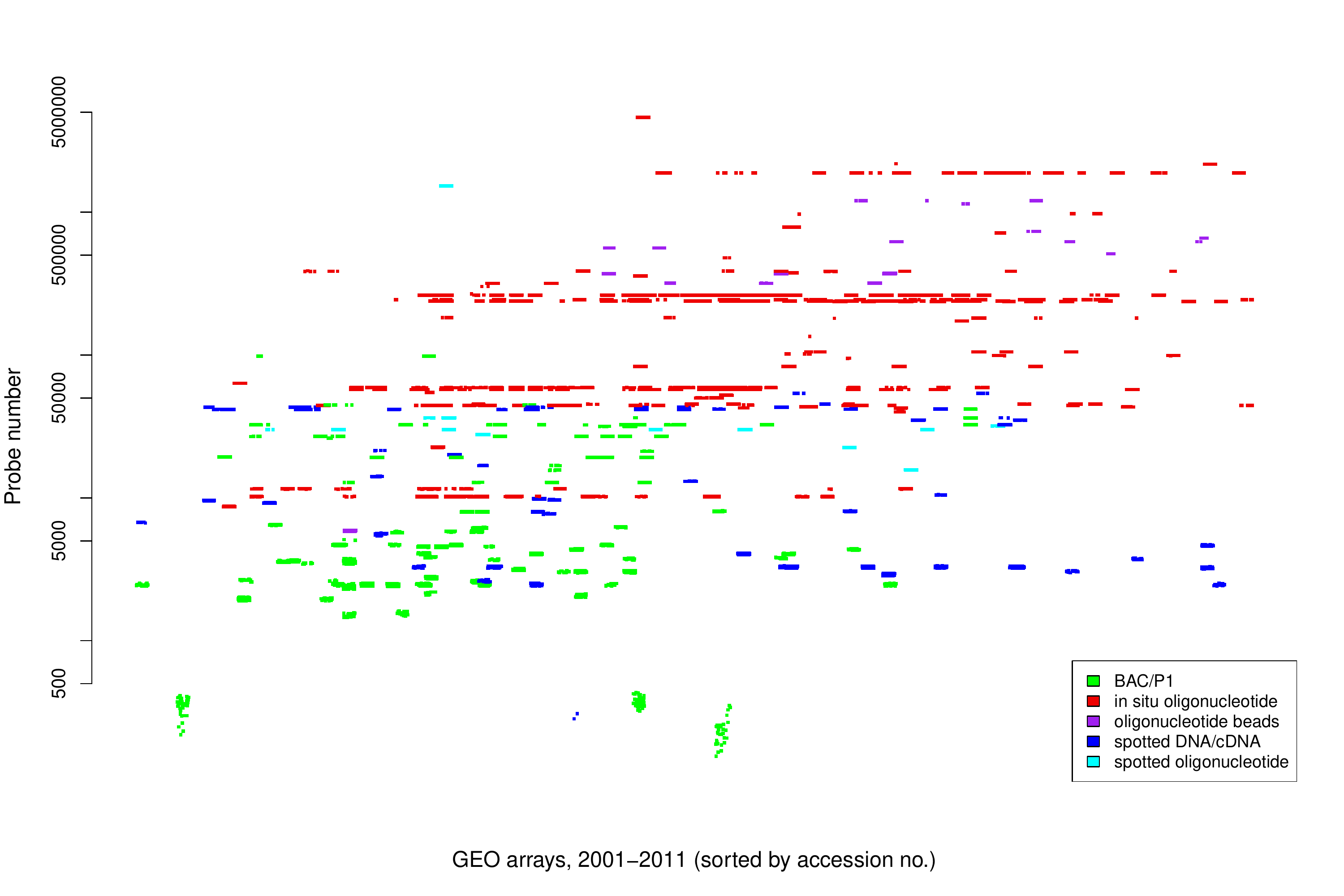}
\caption{Distribution of resolutions and techniques of GEO platforms. Each point represents a genomic array. The Y axis is labeled with probe number in log scale. The X axis denotes the time sequence of array data generation. From left to right are years from 2001 to 2011.}
\label{figure1}
\end{figure}

The most widely available array CGH platforms are either based on large insert clones (BAC/P1 arrays) or based on shorter single-stranded DNA molecules (oligonucleotide arrays), which may or may not include single-nucleotide polymorphism specific probe sequences (SNP arrays). Also, although designed for gene expression profiling, cDNA arrays were used by several laboratories for measuring genomic copy number changes. Although all these platforms are considered suitable for whole genome CNA analysis, their probe densities and other parameters can affect specific features of the analysis results \cite{Baumbusch:2008gc,Curtis:2009fl,Greshock:2007hm,Bengtsson:2009eo}. Table S3 lists the general platform types and corresponding overall numbers of the data registered in arrayMap.

In reviewing the technical platform composition, two related trends become apparent (Figure 1). Originally developed in groups with expertise in molecular cytogenetics and cancer genome analysis, printed large insert clone arrays (BAC/P1) were the first whole genome CNA screening tools with a spatial resolution surpassing that of chromosomal CGH. Other groups re-employed cDNA arrays, developed for expression screening, for genomic hybridizations. However, over the last years one can observe the overwhelming use of various industrially produced oligonucleotide array platforms, which compensate their low single probe fidelity through a probe density at 1-3 orders of magnitude higher than common for BAC/P1 arrays. Another reason for the success of oligonucleotide arrays is the integration of SNP specific probes, which in principle allows to use of the same experiments for genetic association studies and the evaluation of copy number neutral loss of heterozygosity regions \cite{Bignell:2004bu,Heinrichs:2007ed,Carter:2007ck}

\subsection* {Data access and usage scenarios}

Based on our experience from the Progenetix project, a strong emphasis was put on a user friendly data interface. Here, we followed a "dual user type" scenario: Users without bioinformatics background should be able to intuitively visualize core data features as well as to perform standard analysis procedures, while for bioinformaticians the formatted database content should be accessible to use with their analysis tools 
of choice.

\textbf{Query interface.} Data browsing in arrayMap is based on two types of query methods: search by experimental series metadata and search by array features.

In the series query form, users can perform various search options by specifying (i) PubMed ID; (ii) series ID; (iii) platform ID; (iv) platform descriptions and (v) descriptive diagnosis text. For array specific queries, additional features are available: array ID; disease classification (ICD-O 3 code and text) as well as disease locus (code and text) and single or combined regional CNA.

In the results table, associated array information is displayed. A number of links to additional and/or outside  data is provided, according to the information available: the corresponding PubMed entries; the original GEO/AE accession display page for more complete information; the case and publication entries on the Progenetix website for further analysis; and importantly the array specific data visualization page.

\begin{figure}[!ht]
\centering
\includegraphics[width=4in]{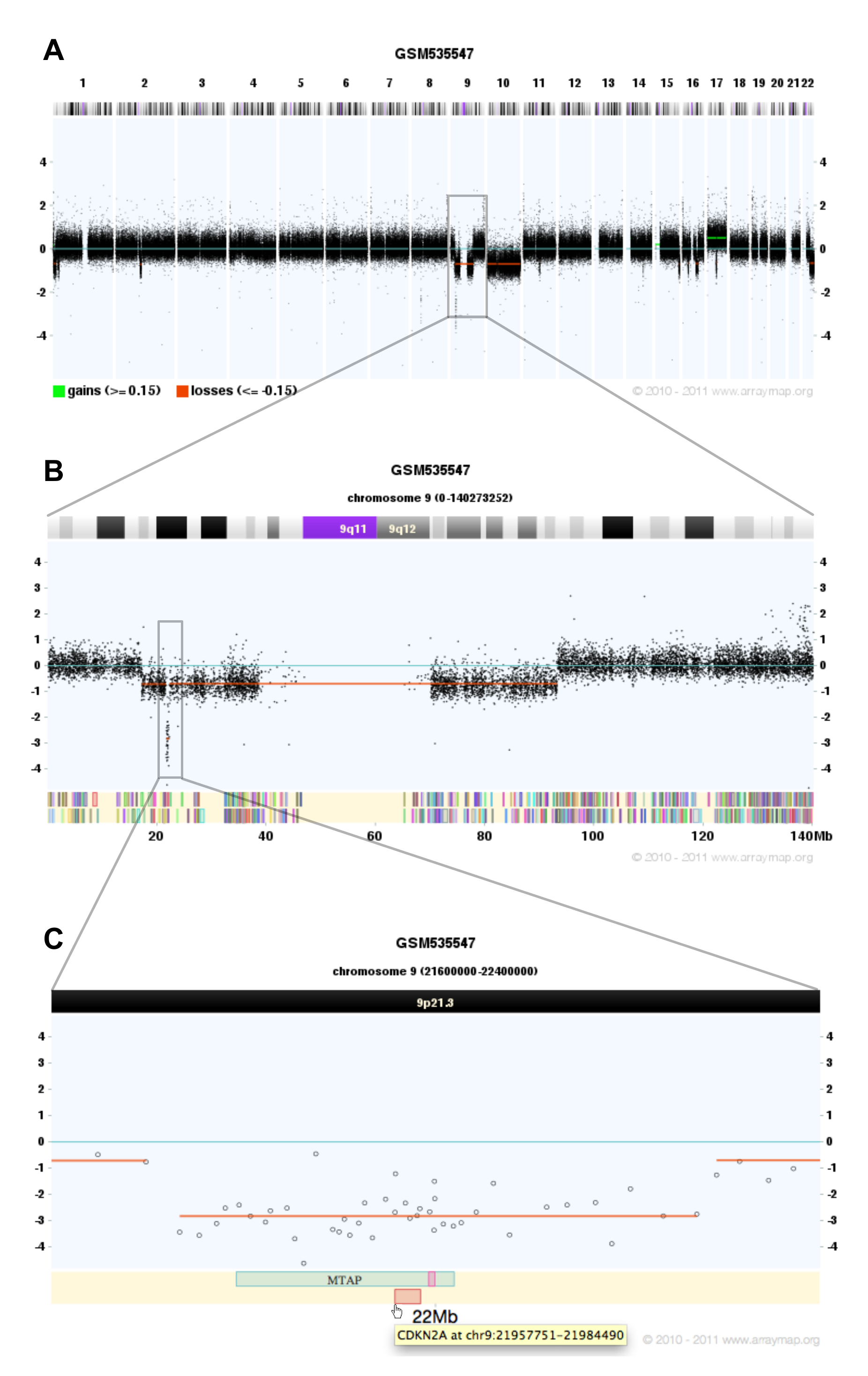}
\caption{Zoom-in visualization of focal CNA. (A) GSM535547 (human high grade glioma, Agilent CGH 244A) shows high quality of probe hybridization signal. CNAs are easy to distinguish. (B) When zoom-in the whole chromosome 9, an approximately 80 MB deletion is displayed, with two breakpoints located in p and q arm respectively. In addition, a small regional deletion in 9p21 is quite clear. Color bars in lower region of the panel represent 848 genes located in chromosome 9. (C) Zoom in the potential homozygously deleted region in 9p21 by specifying the exact region: chr9:21600000-22400000. The zoomed-in plot shows probes, chromosome band and two tumor suppressor genes, MTAP and CDKN2A. Gene name and location will be given while mouse hover. They link to UCSC genome browser with additional information.}
\label{figure2}
\end{figure}

\textbf{Array probe data visualization.} In the array plot interface, original plots of genomic array data sets can be searched and visualized (Supplemental Figure S1). Default threshold parameters which were either  provided with the data or assigned during the initial visualization will be loaded. In single array visualization, the general view of probe distribution and post-thresholding segmentation results are displayed for the whole genome as well as for each individual chromosome. If multiple arrays are retrieved, users can select sample data for downstream analysis procedures. Supplemental Figure S2 shows the screenshot of single array visualization.

Users can segment the raw data values and re-plot the results after revising the following parameters:
\begin{itemize}
  \item Golden path edition, default HG18/NCBI Build 36. This is still the commonly used version of the human reference genome assembly. At the moment, coordinates of probes from all platforms were remapped to HG18. For the near future , we intend to allow for a selection of updated genome editions.
  \item Chromosomes to plot, default 1 to 22. Single or all chromosomes can be selected for re-plotting. To avoid gender bias, most platforms do not contain probes in chromosome X and Y during the design.
  \item Loss/gain thresholds. Cut-offs from which a segment is considered a genomic loss or gain. The optimum thresholds may vary between platforms.
  \item Region size in kb. Sets a filter to remove CNA below (e.g. probable noise) or above (e.g. exclude non-focal CNA) a certain size range.
  \item Minimal probe numbers for segments. This parameter can be used to limit the minimal number of probes required for a segment to be considered (e.g. to remove aberrant segmentation due to probe level noise). Empirical examples would be values of 2-3 for high quality BAC arrays and 6-10 for Affymetrix SNP 6 arrays.
  \item Plot region. Single genomic region to be plotted, overriding the chromosome selection above. When selected, plots with this region will be generated for all current arrays. This is valuable to e.g. display the CNA status and copy number transition points for specific genes of interest (Supplemental Figure S3).
\end{itemize}

\textbf{Zoom-in visualization of focal CNA.} Figure 2 shows the visualization of focal genomic imbalances, e.g. to identify genes of interest targeted by focal CNA. The whole genome view of GSM535547 (human high grade glioma sample analyzed by Agilent Human Genome CGH Microarray 244A) shows a small regional deletion in chromosome 9p21. When plotting the approximate locus of the deletion (specified as chr9:21600000-22400000), genes, probes and chromosome bands in this zoomed in region are shown. Two genes, MTAP and CDKN2A can be seen as being localized in a potential homozygously deleted region. The focal deletion of these known tumor suppressor genes \cite{Lubin:2009de,Krasinskas:2010cn} points to their specific involvement in the glioblastoma sample analyzed here.

\textbf{Querying compound CNA.} The concept of focal CNA detection can be integrated with a global search for arrays containing gene specific regional imbalances. As an example, we demonstrate the search for arrays displaying imbalances in 4 gene loci associated with glioblastoma: EGFR, a transmembrane receptor and proto-oncogene \cite{Smith:2001we}; PTEN, a tumor suppressor gene \cite{Li:1997jk}; ASPM, frequently overexpressed in glioblastoma relative to normal brain tissue \cite{Horvath:2006dm}; and CDKN2A (see above). In the "Search Public Arrays" form, the "Match ..." can be used to specify the genomic regions of those four genes including the expected CNA type: for EGFR (chr7:55054219-55242524:1), PTEN (chr10:89613175-89718511:-1), ASPM (chr1:195319885-195382287:1) and CDKN2A (chr9:21957751-21984490:-1), respectively.

\noindent When executing the query, these regions were matched with the whole database and returned cases which have imbalances overlapping all these regions. When excluding controls and "worst quality" datasets, 303 out of 42421 arrays could be identified matching all four CNA regions. In addition to glioblastoma, several other types of cancer cases were among the results, including e.g. neuroblastomas, breast carcinomas, melanomas and lung carcinomas, which is in accordance with some previous observations \cite{Zhang:2010bt,vanderRhee:2011el,Bourdeaut:2011be,Jin:2011cm}. CNA and associated data of those cases can be processed by online tools for further analysis and visualization (Supplemental Figure S4) or downloaded for offline processing.

\textbf{Copy number profiling of selected cancer entities.} One aim of arrayMap is to allow researchers to conveniently perform aCGH meta-analysis across different platforms. By selecting a single or several cancer entities e.g. based on their ICD entity codes or diagnostic keywords, users are able to generate disease specific CNA frequency profiles or to compare profiles of different cancer types.

As an example, we used ICD-O code 9440/3 (glioblastoma, NOS) to query the database. 1478 arrays from 25 publications were returned and passed to our suite of online analysis tools. Chromosomal ideograms and histograms were generated representing the frequency of copy number aberrations identified over the whole dataset (Figure 3A). In the overall aberration profile, the most common genomic imbalances included whole chromosome 7 gain and chromosome 10 loss, as well as focal gains e.g. on bands 1q21 and 17q21. In our example dataset, a prominent focal deletion hot-spot was centered around 9p21.3 (921 of 1478 arrays, 62.31\%) which had been discussed previously \cite{Wiltshire:2000uj}. The distribution of CNAs over the individual arrays was visualized through a matrix plot (Figure 3B). As additional information to the frequency histograms, this form of visualization facilitates e.g. the detection of CNA patterns among individual arrays as well as the concordance of individual CNAs (e.g. here the arm-level changes in chromosome 7 and 10).

In the matrix plot, clicking on a certain segment would open the related view in the UCSC genome browser \cite{Fujita:2011bf}, for detailed information related to this genomic region (SVG plot only). The plot order of arrays can be re-sorted according to ICD morphology, ICD topography, clinical group or PubMed ID, which can be helpful in associating CNA patterns to external classification categories. For the selected classification criterium (default: ICD morphology), regional CNA frequencies for cases matching the different values will be visualized through a heatmap (Figure 3C); this feature is especially useful when comparing a number of different primary classification criteria.

\subsection* {An overall genomic copy number profile of cancer}

Our high quality core dataset in arrayMap was used to generate an overall cancer copy number aberration profile based on 29,137 arrays (Figure 4).

This data represented 177 cancer types according to ICD-O 3 code, with 59 types among them contained more than 50 arrays. Overall, one of the most common genomic alteration is copy-number gain of chromosome band 8q24, which is found in 30\% of total samples. According to the COSMIC \cite{Forbes:2011cr} database, the most significant cancer gene in this region is MYC. It is a well-documented oncogene codes for a transcription factor that is believed to regulate the expression of 15\% of all genes, including genes involved in cell division, growth, and apoptosis \cite{Gearhart:2007kx,DallaFavera:1982vt}. Other common imbalances observed in at least 25\% of oncogenomic arrays included gains of regions on e.g. 17q21 (29\%), 1q21 (33\%) and loss of regions on e.g. 8p23 (32\%) and 9p21 (25\%), including focal deletions of the CDKN2A/B locus (Figure 2).

While the overall CNA frequency distribution points towards DNA features targeted in multiple entities, this information is insufficient for deriving molecular mechanisms associated with specific cancer types. The genomic heterogeneity of different neoplasias is reflected in the varying patterns of regional CNA frequencies. Based on our core dataset, we have generated a heatmap-style visualization of frequency profiles for all ICD-O entities containing more than 50 arrays (Supplemental Figure S5). The striking patterning of the CNA profiles indicates the non-random occurrence of CNAs, and should be seen as an invitation to explore e.g. CNA similarities shared by separate histopathological entities, as a way to transpose knowledge about pathophysiological mechanisms.

\begin{figure}[ht]
\centering
\includegraphics[width=5in]{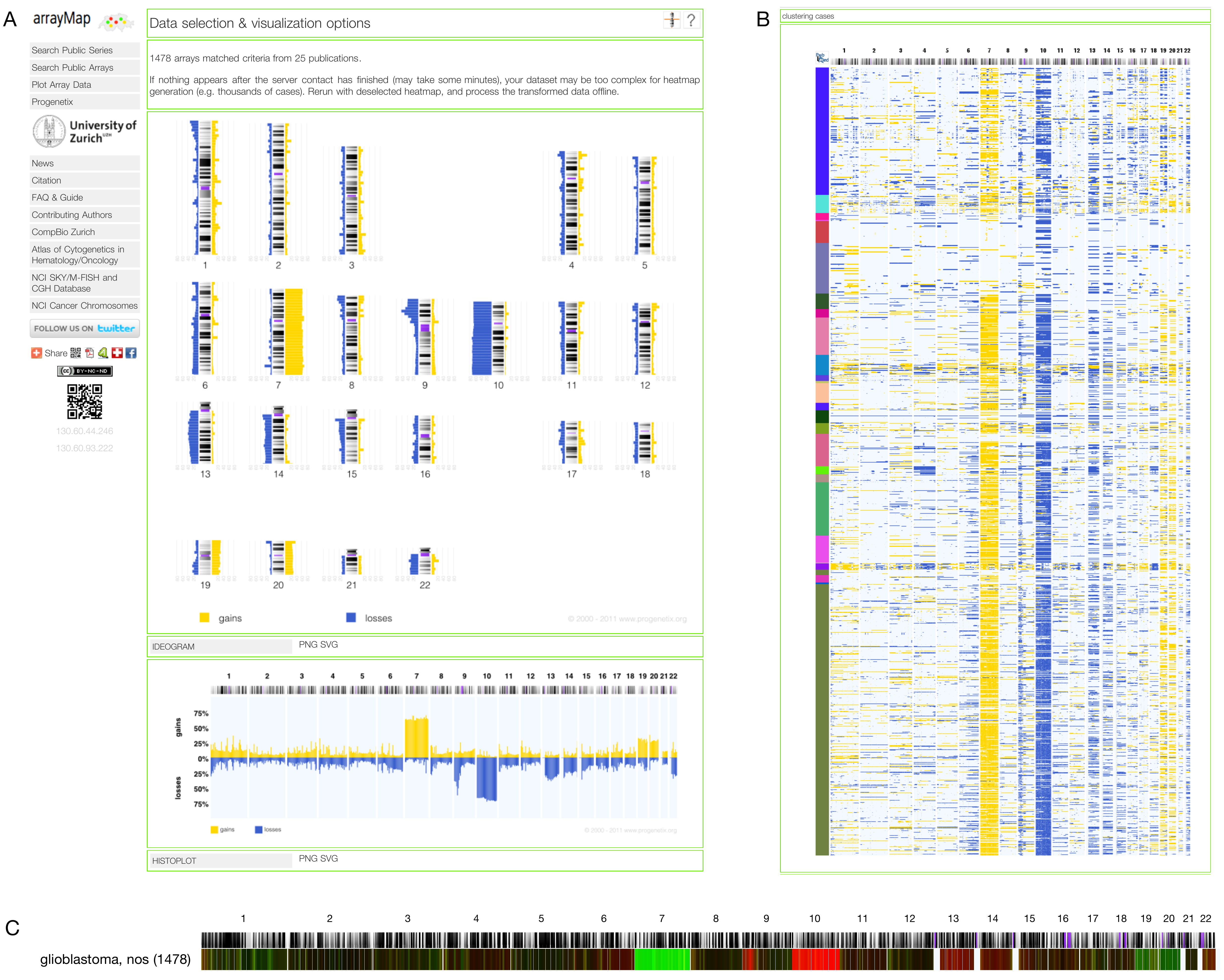}
\caption{Copy number profiling of glioblastoma. (A) Chromosomal ideogram and histogram showing frequency of copy number aberrations. Percentage values corresponding to gains (yellow) and losses (blue) identified over the whole dataset. The most frequent imbalances include gain of chromosome 7 and loss of chromosome 10, 9p21.3. (B) Matrix plot of 1478 glioblastoma cases. The Y axis represents individual samples. The distribution of genomic copy number imbalances reveals the individual aberration patterns of glioblastoma. (C) Heatmap of regional CNA frequencies for 1478 arrays. The intensity of green and red color components correlates to the relative gain and loss frequencies, respectively. If dataset contains cancer subtypes, cancers with similar CNA frequency profiles will be clustered together, such that differences between subtypes will be revealed (e.g. see Figure S4H).}
\label{figure 3}
\end{figure}

\begin{figure}[!h]
\centering
\includegraphics[width=5in]{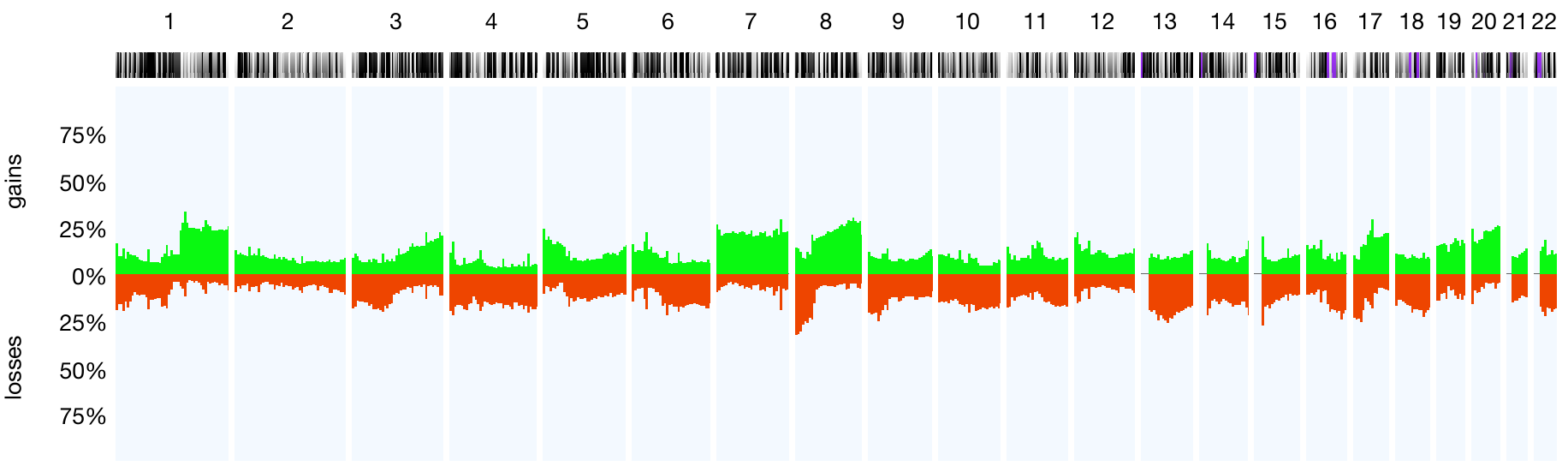}
\caption{The overall cancer copy number aberration profile consisted of 29137 arrays. This plot represents 177 cancer types according to ICD-O 3 code. Percentage values in Y axis corresponding to numbers of gains (green) and losses (red) account for the whole dataset.}
\label{figure 4}
\end{figure}

\section*{Discussion}

\indent arrayMap was developed to facilitate the progress of oncogenomic research. Our aim is to provide high-quality genomic copy number profiles of human tumors, along with a set of tools for accessing and analyzing CNA data. The service has been implemented with a straightforward web interface, including search options for CNA features and clinical annotation data. All assembled datasets are processed into platform independent segmentation and, for the vast majority of arrays, probe level data files, and are presented in consistent formats. Importantly, the direct access to precomputed probe level data plots supports a rapid evaluation of experiments for features of interest. As a curated database using standardized annotation schemes (e.g. ICD classification), arrayMap facilitates the exploration of cancer type specific CNA data, as well as the statistical association of genomic features to clinical parameters.

arrayMap is a dynamic database that is being continuously expanded and improved. We will review the existing and newly published articles to update the database periodically. Over the past decade, we have witnessed a rapidly increasing number of aCGH publications, which gives us sufficient evidences to anticipate that cases in our database will continue to be deposited at a high rate. Although arrayMap is not a user driven repository, we welcome and support users interested in using the site for yet undisclosed data, if they agree on data sharing upon publication.

Although, in contrast to the continuous data from expression analysis, copy number analysis explores discrete value spaces (countable number of DNA copies, for segments defined by genomic base positions), interpretation of the data can vary due to different low level (e.g. signal/background correction) and higher level (e.g. segmentation algorithms, regional or size based filtering) procedures. In that respect, we have to emphasize that the results of our data processing and annotation procedures are open to scrutiny. We encourage a critical review of individual results, and are open for suggestions regarding improved processing procedures for specific platforms.

In this paper, we have provided example scenarios of using arrayMap on different levels, i.e. locus centric and for entity profiling. We believe that systematic analyses will help researchers to discover features which are indiscernible in individual studies, and thus bring tremendous new insights for understanding of disease pathology and inspire the development of new therapeutic approaches. Researchers can integrate arrayMap data with their own analysis efforts, e.g. to increase sample size or for result verification purposes.

We hope that this database will promote further evolution of microarray data meta-analysis. ArrayMap provides access to more than 200 tumor types, which makes it suitable for research across cancer entities. Furthermore, normal sample controls are of vital importance for genomic imbalances studies. ArrayMap includes more than 3000 normal samples from healthy individuals or from normal tissues of cancer patients. These data could be integrated as reference dataset e.g. to account for copy number variation data superimposed on the tumor profiling results.

In the near future, with the continuous accumulation of very high resolution CNA data from genomic arrays and whole genome sequencing experiments, it will become possible to integrate these data into systems biology methods to elucidate effects of genomic instability, and describe the results from more perspectives. Envisioned examples would be e.g. the identification of genes that are involved in metastasis and treatment response; identification of chromosomal breakpoints distribution in cancer; and modeling functional networks in cancer by systems biology approaches.


\section*{Materials and Methods}

\subsection* {Dataset collection}

Raw experimental data from a variety of platforms and repositories were extracted. They were converted to an uniform format which is suited to our reanalysis and visualization system. After a series of parsing procedures, the called copy number data is stored in arrayMap. The flowchart of arrayMap data collection and analysis is as shown in Figure 5. Five main data sources are integrated into arrayMap:\newline

\textbf{GEO/AE.} For extracting appropriate data Series from GEO/AE, two basic criteria have to be fulfilled. First, the raw data has to be from human malignancies analyzed by BAC, cDNA, aCGH or oligonucleotide arrays. Second, the array platform must be genome wide, with the optional omission of the sex chromosomes. Chromosome or region specific arrays were excluded because they were not able to reveal the whole genomic profile of the respective cancer. Associated clinical data was extracted if available.\newline

\textbf{TCGA.} Segmentation data with available clinical information was extracted and incorporated into the database. Due to data sharing restrictions, TCGA data is an exception in that, so far no probe level data is incorporated into arrayMap. This exception was accepted since users will be able to access individual TCGA datasets through the projects web portal at http://tcga-data.nci.nih.gov/tcga/.\newline

\textbf{Publications.} Many aCGH datasets can be found in the text or supplementary files of publications. In order to collect data from publications, we relied on our Progenetix projectÕs setup. Data in Progenetix is manually curated. The collection strategies are:
\begin{itemize}
  \item literature mining using complex search parameters through PubMed
  \item identification of called aCGH data, in GP annotation or tabular format (article, supplementary tables)
  \item evaluation of supplementary files for probe specific data tables
  \item follow-up on article links outs, to repository entries or referenced datasets\newline
\end{itemize}
\textbf{User submission.} User submitted data was provided in a number of formats which were converted to the standard format as described. Although we accept and support private datasets, we insist on integration of  at least the genomic and core clinical data (e.g. disease classifiers) upon publication of the datasets analysis results.

\begin{figure}[!ht]
\centering
\includegraphics[width=4in]{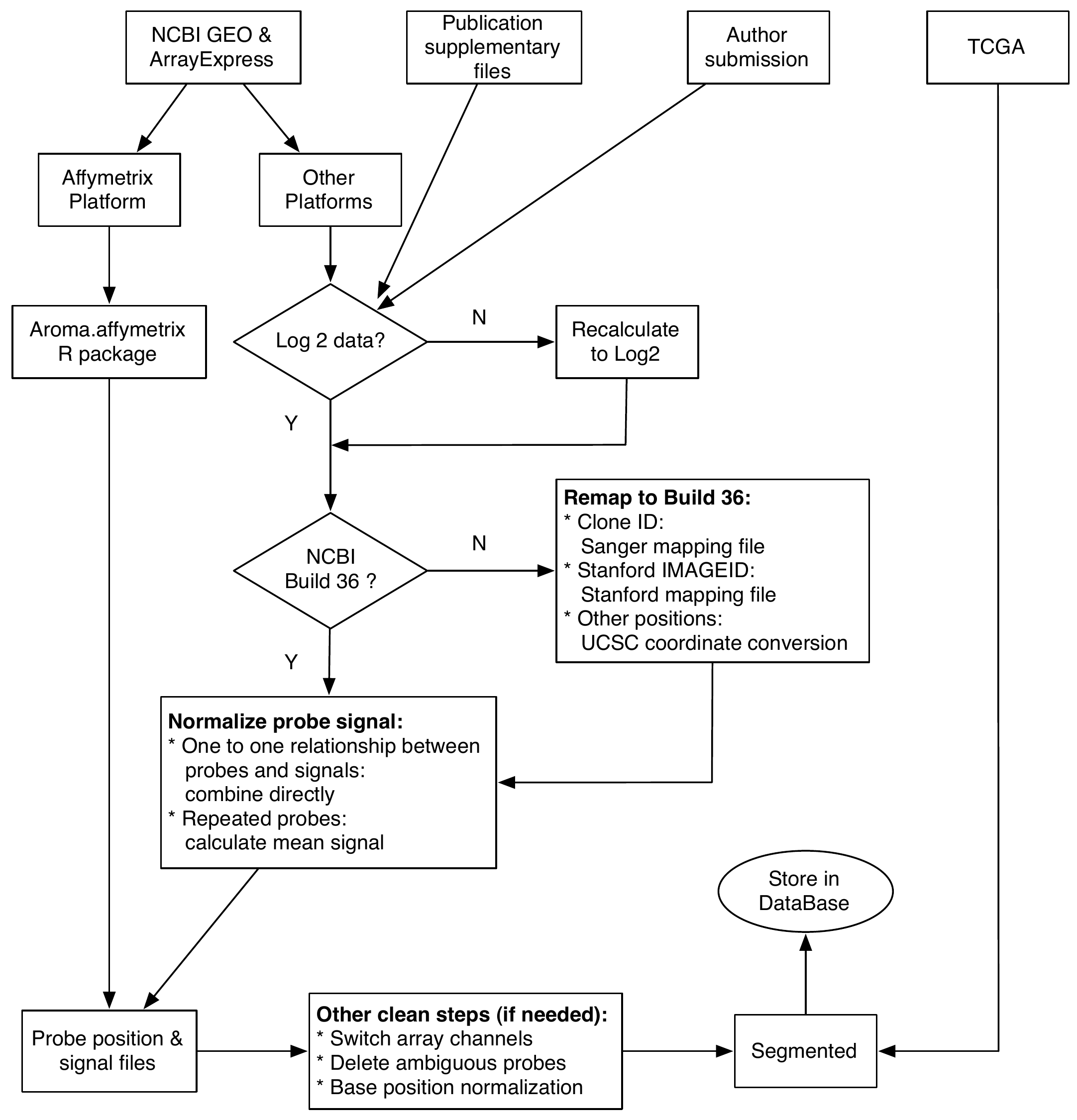}
\caption{The flowchart of arrayMap data collection and analysis procedures. Publicly available raw data or segmented data was collected from respective data sources. Files were reprocessed by distinct procedures according to different data types. All kinds of probe signals were converted to log2 value. Probe coordinates were remapped to the most commonly used human reference genome assembly (NCBI Build 36/hg18). At last, all information was converted to uniform format and stored in arrayMap, which is accessed by the web application.}
\label{figure5}
\end{figure}

\subsection*{Dataset analysis}

\textbf{Probe remapping.} A pipeline has been generated for determining the genomic positions for the tens to hundreds of thousands array probes with reference to a common genome Golden Path edition. For each array platform, the genome positions of probes were remapped to the current commonly used version of the human reference genome assembly (NCBI Build 36.1/hg18). Specific mapping procedures were employed for different types of probes. BAC clones were firstly remapped according to the clone sets information of Sanger/DECIPHER database \cite{Firth:2009ck}. If the probe position was not available, the UCSC Genome annotation database \cite{Fujita:2011bf} (release hg18) was used for compensation. After these two steps, a mean of 98\% of the BAC clones were remapped. For IMAGE clone sets, only the UCSC Genome annotation database was used. The average remapping rate of IMAGE clones was 91\%. Affymetrix raw CEL data files were analyzed based on hg18 library files, namely the output segments have hg18 coordinates. The summary of the percentage of mapped probes is given in Table 3. The mapping details for each platform can be found in the supplementary data  (Table S4).\newline

\begin{table}[!ht]
\caption{
\bf{Percentage of remapped probes according to platform types}}
\setlength{\tabcolsep}{11pt}
\setlength{\extrarowheight}{6pt}
\begin{tabular}{lllll}
\hline
Platform type&Average mapping rate&Number of arrays&Number of GPLs\\
\hline
Original HG18 (Build 36)&NA&1583&40\\
in situ oligonucleotide&99\%&21678&55\\
BAC/P1&98\%&5464&55\\
spotted DNA/cDNA&91\%&2365&82\\
\hline
\end{tabular}
\label{tab:label}
 \end{table}

\textbf{Probe signal normalization.} The array data available was given in a variety of formats, most frequently as log2 ratio of probe hybridization intensity. In order to make data from different platforms directly comparable, all other types of normalized values were converted to log2. For dye swap experiments, reference/tumor intensity ratios data was "reversed" representing a tumor/reference value. For some two-color arrays for which only raw signal intensity were provided, the normalized log2 ratio for each probe was calculated by:

r = log2((T$_{s}$ - T$_{b}$) / (R$_{s}$ - R$_{b}$))

where T$_{s}$ and T$_{b}$ represent tumor sample intensity and tumor channel background intensity respectively, and R$_{s}$ and R$_{b}$ represent reference sample intensity and reference channel background intensity respectively.

If multiple instances of the same clone exist, the average signal intensity of the certain clone was considered.\newline

\textbf{Affymetrix genotyping arrays.} For the widely used Affymetrix GenomeWide SNP arrays, raw CEL files were downloaded and underwent a massive re-analysis using the R package aroma.affymetrix \cite{Bengtsson:2008tga} with the CRMAv.2 method \cite{Bengtsson:2009ho}. During the processing step, approximately 50 normal sample arrays were employed as a reference set for each array type to reduce the noise level. Normal tissue arrays from different labs were extracted and used to build the reference dataset. In order to obtain high quality arrays, we excluded arrays which contain segments greater than 3 mega-bases, since copy number variations are always smaller than 3 mega-bases. The list of normal tissue reference arrays is giving in Table S5 in Supplementary data.\newline

\textbf{Quality control.} In our review of array data deposited in GEO or collected from publication supplements we encountered a large number of individual data sets with insufficient or limited probe quality. Also, for samples of unprocessed raw data (e.g. Affymetrix CEL files), we found that QC measures reported previously (e.g. call rate \cite{Laurie:2010cs}, NUSE \cite{F:2005vha}, RLE \cite{F:2005vha}) only had a limited accuracy for detection of arrays with inadequate probe level data.
Currently, the most viable strategy for quality assessment of processed, heterogeneous copy number arrays is the visual inspection of probe plotting and segmentation results through an experienced researcher. 
For the first arrayMap edition we generated a quality classification system, which contains a total of 4 categories based on inspections of genome-wide array plots:
\begin{itemize}
  \item Excellent. Probe signal distribution is significantly different between normal regions and imbalance regions. Signal baseline is distinct and unique, making segmentation threshold realistic appearing. Chromosomal changes are pretty clear.
  \item Good. In general good quality. Probe signal may contain some noise, but tolerable. Chromosomal changes are distinguishable.
  \item Hypersegmented. Serrated distribution of probe signal intensities, causing dozens of separate peaks and discontinuous segments. Chromosomal changes are always up to several hundreds and smaller than 5 mega-bases.
  \item Noisy. Probe signal intensities are highly scattered, but well-distributed, with high standard deviation, resulting in the inability to differentiate copy number changes.
\end{itemize}
Depending on the intended research purpose this basic classification system can be used for a pre-analysis triage of copy number data. Applying stringent review criteria we identified a core dataset with "excellent" quality arrays accounting for approximately 60 percent of total arrays.

We are currently working on a platform independent quality assessment system for genomic arrays, which will be implemented in future versions of the arrayMap resource.\newline

\textbf{Associated data.} For arrayMap, data is stored with separate datasets for each array. This is in contrast to the Progenetix database, for which technical replicates where available are combined into case specific CNA profiles. In arrayMap, technical replicates are assigned an identical case identifier to facilitate downstream statistical procedures including e.g. clinical data correlations. The assignment of the correct diagnostic entity to each sample is an essential step in generating a binding between genomic and associated data points. At the same time, to ensure annotation consistency and make the retrieval process more efficient, for all CNA profiles the following data points were manually collected from GEO/ArrayExpress and published papers if available.
\begin{itemize}
  \item Descriptive diagnostic text, as available through the original source
  \item Diagnostic classification according to the International Classification of Diseases in Oncology (ICDO 3, morphology with code)
  \item Tumor locus according to ICD (ICD topography with code)
  \item Source of material (e.g. primary tumor, cell line, metastasis)
  \item Clinical parameters where available, including age, gender, grade, clinical stage (TNM coded), recurrence/progression, time to recurrence/progression, death and followup\newline
\end{itemize}

\textbf{Web Server} An online interface of arrayMap database was created using Perl common gateway interface (CGI) and R scripts running on a Mac OS X Server. Data is stored as flat files in the JSON format. Precomputed array plots are stored in SVG and PNG versions. The online release of the service has been optimized to be compatible with major browsers supporting current web standards (CSS2, HTML5, XML with inline SVG; e.g. Safari \textgreater= 3.0, Firefox \textgreater= 3.0, InternetExplorer \textgreater= 9, Crome) with limited fallback support. Dynamic graphics provided in the array plot module were implemented by technologies including XML/XHTML, JavaScript, SVG and HTML5 Canvas.

\section*{Acknowledgments}
We thank Christian von Mering, Homayoun Bagheri and Nuria Lopez-Bigas for helpful discussions.

\section*{Author Contributions}
Conceived and designed the project: MB. Performed the experiments: HC NK. Analyzed the data: HC NK. Contributed reagents/materials/analysis tools: MB HC. Wrote the paper: HC MB.

\bibliography{Cai_Kumar_Baudis_arrayMap}

\begin{thebibliography}{10}
\providecommand{\url}[1]{\texttt{#1}}
\providecommand{\urlprefix}{URL }
\expandafter\ifx\csname urlstyle\endcsname\relax
  \providecommand{\doi}[1]{doi:\discretionary{}{}{}#1}\else
  \providecommand{\doi}{doi:\discretionary{}{}{}\begingroup
  \urlstyle{rm}\Url}\fi
\providecommand{\bibAnnoteFile}[1]{%
  \IfFileExists{#1}{\begin{quotation}\noindent\textsc{Key:} #1\\
  \textsc{Annotation:}\ \input{#1}\end{quotation}}{}}
\providecommand{\bibAnnote}[2]{%
  \begin{quotation}\noindent\textsc{Key:} #1\\
  \textsc{Annotation:}\ #2\end{quotation}}
\providecommand{\eprint}[2][]{\url{#2}}

\bibitem{Stallings:2007fh}
Stallings RL (2007) {Are chromosomal imbalances important in cancer?}
\newblock Trends in genetics : TIG 23: 278--283.
\bibAnnoteFile{Stallings:2007fh}

\bibitem{Myllykangas:2006hv}
Myllykangas S, Himberg J, B{\"o}hling T, Nagy B, Hollm{\'e}n J, et~al. (2006)
  {DNA copy number amplification profiling of human neoplasms.}
\newblock Oncogene 25: 7324--7332.
\bibAnnoteFile{Myllykangas:2006hv}

\bibitem{Weir:2007km}
Weir BA, Woo MS, Getz G, Perner S, Ding L, et~al. (2007) {Characterizing the
  cancer genome in lung adenocarcinoma.}
\newblock Nature 450: 893--898.
\bibAnnoteFile{Weir:2007km}

\bibitem{Wiedemeyer:2008kl}
Wiedemeyer R, Brennan C, Heffernan TP, Xiao Y, Mahoney J, et~al. (2008)
  {Feedback circuit among INK4 tumor suppressors constrains human glioblastoma
  development.}
\newblock Cancer cell 13: 355--364.
\bibAnnoteFile{Wiedemeyer:2008kl}

\bibitem{Mullighan:2007jv}
Mullighan CG, Goorha S, Radtke I, Miller CB, Coustan-Smith E, et~al. (2007)
  {Genome-wide analysis of genetic alterations in acute lymphoblastic
  leukaemia.}
\newblock Nature 446: 758--764.
\bibAnnoteFile{Mullighan:2007jv}

\bibitem{CancerGenomeAtlasResearchNetwork:2008gr}
{Cancer Genome Atlas Research Network} (2008) {Comprehensive genomic
  characterization defines human glioblastoma genes and core pathways.}
\newblock Nature 455: 1061--1068.
\bibAnnoteFile{CancerGenomeAtlasResearchNetwork:2008gr}

\bibitem{Kan:2010fo}
Kan Z, Jaiswal BS, Stinson J, Janakiraman V, Bhatt D, et~al. (2010) {Diverse
  somatic mutation patterns and pathway alterations in human cancers.}
\newblock Nature 466: 869--873.
\bibAnnoteFile{Kan:2010fo}

\bibitem{Bergamaschi:2006fj}
Bergamaschi A, Kim YH, Wang P, S{\o}rlie T, Hernandez-Boussard T, et~al. (2006)
  {Distinct patterns of DNA copy number alteration are associated with
  different clinicopathological features and gene-expression subtypes of breast
  cancer.}
\newblock Genes, chromosomes {\&} cancer 45: 1033--1040.
\bibAnnoteFile{Bergamaschi:2006fj}

\bibitem{Hu:2009ez}
Hu X, Stern HM, Ge L, O'Brien C, Haydu L, et~al. (2009) {Genetic alterations
  and oncogenic pathways associated with breast cancer subtypes.}
\newblock Molecular cancer research : MCR 7: 511--522.
\bibAnnoteFile{Hu:2009ez}

\bibitem{Kallioniemi:1992ud}
Kallioniemi A, Kallioniemi OP, Sudar D, Rutovitz D, Gray JW, et~al. (1992)
  {Comparative genomic hybridization for molecular cytogenetic analysis of
  solid tumors.}
\newblock Science (New York, NY) 258: 818--821.
\bibAnnoteFile{Kallioniemi:1992ud}

\bibitem{Pollack:1999by}
Pollack JR, Perou CM, Alizadeh AA, Eisen MB, Pergamenschikov A, et~al. (1999)
  {Genome-wide analysis of DNA copy-number changes using cDNA microarrays.}
\newblock Nature genetics 23: 41--46.
\bibAnnoteFile{Pollack:1999by}

\bibitem{Bignell:2004bu}
Bignell GR, Huang J, Greshock J, Watt S, Butler A, et~al. (2004)
  {High-resolution analysis of DNA copy number using oligonucleotide
  microarrays.}
\newblock Genome research 14: 287--295.
\bibAnnoteFile{Bignell:2004bu}

\bibitem{Baudis:2007er}
Baudis M (2007) {Genomic imbalances in 5918 malignant epithelial tumors: an
  explorative meta-analysis of chromosomal CGH data.}
\newblock BMC cancer 7: 226.
\bibAnnoteFile{Baudis:2007er}

\bibitem{Alloza:2011be}
Alloza E, Al-Shahrour F, Cigudosa JC, Dopazo J (2011) {A large scale survey
  reveals that chromosomal copy-number alterations significantly affect gene
  modules involved in cancer initiation and progression.}
\newblock BMC medical genomics 4: 37.
\bibAnnoteFile{Alloza:2011be}

\bibitem{Barrett:2011gr}
Barrett T, Troup DB, Wilhite SE, Ledoux P, Evangelista C, et~al. (2011) {NCBI
  GEO: archive for functional genomics data sets--10 years on.}
\newblock Nucleic acids research 39: D1005--10.
\bibAnnoteFile{Barrett:2011gr}

\bibitem{Parkinson:2010fj}
Parkinson H, Sarkans U, Kolesnikov N, Abeygunawardena N, Burdett T, et~al.
  (2010) {ArrayExpress update--an archive of microarray and high-throughput
  sequencing-based functional genomics experiments}.
\newblock Nucleic acids research 39: D1002--D1004.
\bibAnnoteFile{Parkinson:2010fj}

\bibitem{Scheinin:2008ff}
Scheinin I, Myllykangas S, Borze I, B{\"o}hling T, Knuutila S, et~al. (2008)
  {CanGEM: mining gene copy number changes in cancer.}
\newblock Nucleic acids research 36: D830--5.
\bibAnnoteFile{Scheinin:2008ff}

\bibitem{Cao:2011ft}
Cao Q, Zhou M, Wang X, Meyer CA, Zhang Y, et~al. (2011) {CaSNP: a database for
  interrogating copy number alterations of cancer genome from SNP array data.}
\newblock Nucleic acids research 39: D968--74.
\bibAnnoteFile{Cao:2011ft}

\bibitem{Baudis:2001uaa}
Baudis M, Cleary ML (2001) {Progenetix.net: an online repository for molecular
  cytogenetic aberration data.}
\newblock Bioinformatics (Oxford, England) 17: 1228--1229.
\bibAnnoteFile{Baudis:2001uaa}

\bibitem{Baumbusch:2008gc}
Baumbusch LO, Aar{\o}e J, Johansen FE, Hicks J, Sun H, et~al. (2008)
  {Comparison of the Agilent, ROMA/NimbleGen and Illumina platforms for
  classification of copy number alterations in human breast tumors.}
\newblock BMC genomics 9: 379.
\bibAnnoteFile{Baumbusch:2008gc}

\bibitem{Curtis:2009fl}
Curtis C, Lynch AG, Dunning MJ, Spiteri I, Marioni JC, et~al. (2009) {The
  pitfalls of platform comparison: DNA copy number array technologies
  assessed.}
\newblock BMC genomics 10: 588.
\bibAnnoteFile{Curtis:2009fl}

\bibitem{Greshock:2007hm}
Greshock J, Feng B, Nogueira C, Ivanova E, Perna I, et~al. (2007) {A comparison
  of DNA copy number profiling platforms.}
\newblock Cancer research 67: 10173--10180.
\bibAnnoteFile{Greshock:2007hm}

\bibitem{Bengtsson:2009eo}
Bengtsson H, Ray A, Spellman P, Speed TP (2009) {A single-sample method for
  normalizing and combining full-resolution copy numbers from multiple
  platforms, labs and analysis methods.}
\newblock Bioinformatics (Oxford, England) 25: 861--867.
\bibAnnoteFile{Bengtsson:2009eo}

\bibitem{Heinrichs:2007ed}
Heinrichs S, Look T (2007) {Identification of structural aberrations in cancer
  by SNP array analysis}.
\newblock Genome biology : 1--5.
\bibAnnoteFile{Heinrichs:2007ed}

\bibitem{Carter:2007ck}
Carter NP (2007) {Methods and strategies for analyzing copy number variation
  using DNA microarrays}.
\newblock Nature genetics 39: S16--S21.
\bibAnnoteFile{Carter:2007ck}

\bibitem{Lubin:2009de}
Lubin M, Lubin A (2009) {Selective killing of tumors deficient in
  methylthioadenosine phosphorylase: a novel strategy.}
\newblock PloS one 4: e5735.
\bibAnnoteFile{Lubin:2009de}

\bibitem{Krasinskas:2010cn}
Krasinskas AM, Bartlett DL, Cieply K, Dacic S (2010) {CDKN2A and MTAP deletions
  in peritoneal mesotheliomas are correlated with loss of p16 protein
  expression and poor survival.}
\newblock Modern pathology : an official journal of the United States and
  Canadian Academy of Pathology, Inc 23: 531--538.
\bibAnnoteFile{Krasinskas:2010cn}

\bibitem{Smith:2001we}
Smith JS, Tachibana I, Passe SM, Huntley BK, Borell TJ, et~al. (2001) {PTEN
  mutation, EGFR amplification, and outcome in patients with anaplastic
  astrocytoma and glioblastoma multiforme.}
\newblock Journal of the National Cancer Institute 93: 1246--1256.
\bibAnnoteFile{Smith:2001we}

\bibitem{Li:1997jk}
Li J (1997) {PTEN, a Putative Protein Tyrosine Phosphatase Gene Mutated in
  Human Brain, Breast, and Prostate Cancer}.
\newblock Science (New York, NY) 275: 1943--1947.
\bibAnnoteFile{Li:1997jk}

\bibitem{Horvath:2006dm}
Horvath S, Zhang B, Carlson M, Lu KV, Zhu S, et~al. (2006) {Analysis of
  oncogenic signaling networks in glioblastoma identifies ASPM as a molecular
  target.}
\newblock Proceedings of the National Academy of Sciences of the United States
  of America 103: 17402--17407.
\bibAnnoteFile{Horvath:2006dm}

\bibitem{Zhang:2010bt}
Zhang W, Zhu J, Bai J, Jiang H, Liu F, et~al. (2010) {Comparison of the
  inhibitory effects of three transcriptional variants of CDKN2A in human lung
  cancer cell line A549.}
\newblock Journal of experimental {\&} clinical cancer research : CR 29: 74.
\bibAnnoteFile{Zhang:2010bt}

\bibitem{vanderRhee:2011el}
van~der Rhee JI, Krijnen P, Gruis NA, de~Snoo FA, Vasen HFA, et~al. (2011)
  {Clinical and histologic characteristics of malignant melanoma in families
  with a germline mutation in CDKN2A.}
\newblock Journal of the American Academy of Dermatology .
\bibAnnoteFile{vanderRhee:2011el}

\bibitem{Bourdeaut:2011be}
Bourdeaut F, Isidor B, Ferrand S, Thomas C, Moreau A, et~al. (2011) {Homozygous
  PTEN deletion in neuroblastoma arising in a child with Cowden syndrome.}
\newblock American journal of medical genetics Part A 155: 1763--1766.
\bibAnnoteFile{Bourdeaut:2011be}

\bibitem{Jin:2011cm}
Jin K, Kong X, Shah T, Penet MF, Wildes F, et~al. (2011) {Breast Cancer Special
  Feature: The HOXB7 protein renders breast cancer cells resistant to tamoxifen
  through activation of the EGFR pathway.}
\newblock Proceedings of the National Academy of Sciences of the United States
  of America .
\bibAnnoteFile{Jin:2011cm}

\bibitem{Wiltshire:2000uj}
Wiltshire RN, Rasheed BK, Friedman HS, Friedman AH, Bigner SH (2000)
  {Comparative genetic patterns of glioblastoma multiforme: potential
  diagnostic tool for tumor classification.}
\newblock Neuro-oncology 2: 164--173.
\bibAnnoteFile{Wiltshire:2000uj}

\bibitem{Fujita:2011bf}
Fujita PA, Rhead B, Zweig AS, Hinrichs AS, Karolchik D, et~al. (2011) {The UCSC
  Genome Browser database: update 2011.}
\newblock Nucleic acids research 39: D876--82.
\bibAnnoteFile{Fujita:2011bf}

\bibitem{Forbes:2011cr}
Forbes SA, Bindal N, Bamford S, Cole C, Kok CY, et~al. (2011) {COSMIC: mining
  complete cancer genomes in the Catalogue of Somatic Mutations in Cancer.}
\newblock Nucleic acids research 39: D945--50.
\bibAnnoteFile{Forbes:2011cr}

\bibitem{Gearhart:2007kx}
Gearhart J, Pashos EE, Prasad MK (2007) {Pluripotency redux--advances in
  stem-cell research.}
\newblock The New England journal of medicine 357: 1469--1472.
\bibAnnoteFile{Gearhart:2007kx}

\bibitem{DallaFavera:1982vt}
Dalla-Favera R, Bregni M, Erikson J, Patterson D, Gallo RC, et~al. (1982)
  {Human c-myc onc gene is located on the region of chromosome 8 that is
  translocated in Burkitt lymphoma cells}.
\newblock Proceedings of the National Academy of Sciences of the United States
  of America Vol. 79: 7824--7827.
\bibAnnoteFile{DallaFavera:1982vt}

\bibitem{Firth:2009ck}
Firth HV, Richards SM, Bevan AP, Clayton S, Corpas M, et~al. (2009) {DECIPHER:
  Database of Chromosomal Imbalance and Phenotype in Humans Using Ensembl
  Resources}.
\newblock The American Journal of Human Genetics 84: 524--533.
\bibAnnoteFile{Firth:2009ck}

\bibitem{Bengtsson:2008tga}
Bengtsson H, Simpson K, Bullard J, Hansen K (2008) {aroma.affymetrix: A genetic
  framework in R for analyzing small to very large Affymetrix data sets in
  bounded memory}.
\newblock Tech Report {\#}745 Department of Statistics, University of
  California, Berkeley .
\bibAnnoteFile{Bengtsson:2008tga}

\bibitem{Bengtsson:2009ho}
Bengtsson H, Wirapati P, Speed TP (2009) {A single-array preprocessing method
  for estimating full-resolution raw copy numbers from all Affymetrix
  genotyping arrays including GenomeWideSNP 5 {\&} 6}.
\newblock Bioinformatics (Oxford, England) 25: 2149--2156.
\bibAnnoteFile{Bengtsson:2009ho}

\bibitem{Laurie:2010cs}
Laurie CC, Doheny KF, Mirel DB, Pugh EW, Bierut LJ, et~al. (2010) {Quality
  control and quality assurance in genotypic data for genome-wide association
  studies}.
\newblock Genetic Epidemiology 34: 591--602.
\bibAnnoteFile{Laurie:2010cs}

\bibitem{F:2005vha}
F C, AL A, SA K, TP S, VL SM (2005) {NUSE and RLE: Quality assessment of
  oligonucleotide microarray data to quantify systemic variation}.
\newblock 2005 Meeting of the Federation of Clinical Immunology Societies
  Boston, MA .
\bibAnnoteFile{F:2005vha}

\end{thebibliography}

\section*{Supporting information}

{\bf Figure S1. Array data sets visualization.} Original plots and optimized parameters for GSE21530 which contains 8 intimal sarcoma samples hybridized on Agilent CGH Microarray 244A platform. The normalized probe signal log2 ratios and post-thresholding segmentation results for each array are intuitively displayed. Genomic alterations are represented by horizontal green (gain) and red (loss) lines. Alterations defined here as regions with log2 ratio \textgreater0.15 or \textless-0.15. Simplified schemas of CNAs link to UCSC genome browser for further review.\\\\
{\bf Figure S2. Screenshot of single array visualization.} ArrayMap plots for GSM630977 (acute myelogenous leukemia). Besides the whole genome view, subviews of each chromosome are displayed as well. From these plots, different kinds of genetic variation events are clearly revealed, e.g. massive genomic rearrangement in chromosome 6; arm-level gain of chromosome 8q and 3MB focal change around 1p31.3. Through the "Plot Array Data" interface, users can segment the raw data values and re-plot the results with customized parameters.\\\\
{\bf Figure S3. Plot single genomic region.} In the ÒPlot Array DataÓ interface, input the precise location (chr5:1100000-1400000) in ÒPlot RegionÓ field. Plots with this region were generated for all 8 arrays in the current series (GSE21530). In this region, there are 5 genes which are shown schematically as colored boxes. CNA status and copy number transition points for these genes are displayed.\\\\
{\bf Figure S4. Compound CNA query.} (A) Four gene loci associated with glioblastoma (EGFR, PTEN, ASPM and CDKN2A) were inserted into "Match regions" field. 303 out of 42421 arrays were returned. (B) Classification information of these 303 arrays were displayed and can be selected for the following analysis. (C) Statistical and plot parameters can be customized. Associated data was processed by online tools, and returned results included: (D) Chromosomal ideogram and (E) histogram, show frequency of copy number aberrations; (F) Matrix plot reveals the aberration pattern of selected arrays; (G) Array classification tree generated by hierarchical Ward clustering, arrays with similar frequency of CNA are part of the tree branch. (H) Heatmap of CNA frequencies clustered by clinical group.\\\\
{\bf Figure S5. Heatmap of frequency profiles for 59 cancer types.} Heatmap visualization of frequency profiles for all ICD-O entities containing more than 50 arrays in our core dataset. Region specific gain/loss frequencies were mapped to 1MB intervals. The intensity of colors (green: gains; losses: red) corresponds to the relative frequency of CNAs for each interval.\\\\
{\bf Table S1: Entities extracted from NCBI GEO and EBI ArrayExpress}\\
{\bf Table S2: Cancer entities grouped by ICD-O code}\\
{\bf Table S3: Platform type distribution in arrayMap}\\
{\bf Table S4: Probe remapping rate for platforms}\\
{\bf Table S5: Normal tissue reference arrays for Affymetrix platforms}\\

\end{document}